\documentclass[pdflatex,oneside,twocolumn,sn-mathphys-num]{sn-jnl}



\usepackage{graphicx}%
\usepackage{multirow}%
\usepackage{amsmath,amssymb,amsfonts}%
\usepackage{amsthm}%
\usepackage{mathrsfs}%
\usepackage[title]{appendix}%
\usepackage{xcolor}%
\usepackage{textcomp}%
\usepackage{manyfoot}%
\usepackage{booktabs}%
\usepackage{algorithm}%
\usepackage{algorithmicx}%
\usepackage{algpseudocode}%
\usepackage{listings}%

\theoremstyle{thmstyleone}%
\theoremstyle{thmstyletwo}%

\theoremstyle{thmstylethree}%

\begin{document}

\title[Article Title]{Integrated Sensing and Communication: Towards Multifunctional Perceptive Network}

\author[1]{\fnm{} \sur{Yuanhao Cui}}\email{yuanhao.cui@bupt.edu.cn}

\author*[2]{\fnm{} \sur{Jiali Nie}}\email{niejl@seu.edu.cn}

\author*[2]{\fnm{} \sur{Fan Liu}}\email{fan.liu@seu.edu.cn}

\author[3]{\fnm{} \sur{Weijie Yuan}}\email{yuanwj@sustech.edu.cn}

\author[1]{\fnm{} \sur{Zhiyong Feng}}\email{fengzy@bupt.edu.cn}

\author[1]{\fnm{} \sur{Xiaojun Jing}}\email{jxiaojun@bupt.edu.cn}

\author[4]{\fnm{} \sur{Yulin Liu}}\email{liuyulin@cnnic.cn}

\author*[5]{\fnm{} \sur{Jie Xu}}\email{xujie@cuhk.edu.cn}

\author[6]{\fnm{} \sur{Christos Masouros}}\email{c.masouros@ucl.ac.uk}


\author[5]{\fnm{} \sur{Shuguang Cui}}\email{shuguangcui@cuhk.edu.cn}

\affil[1]{\orgdiv{School of Information and Communication Engineering}, \orgname{Beijing University of Posts and Telecommunications}, \orgaddress{\city{Beijing}, \postcode{100876}, \country{China}}}

\affil*[2]{\orgdiv{National Mobile Communications Research Laboratory}, \orgname{Southeast University}, \orgaddress{\city{Nanjing}, \postcode{211189}, \country{Chforina}}}

\affil[3]{\orgname{Southern University of Science and Technology}, \orgaddress{\city{Shenzhen}, \postcode{518055}, \country{China}}}

\affil[4]{\orgname{Beijing China Internet Network Information Center}, \orgaddress{\city{Beijing}, \postcode{211189}, \country{China}}}

\affil[5]{\orgname{The Chinese University of Hong Kong, Shenzhen}, \orgaddress{\city{Shenzhen}, \postcode{518172}, \country{China}}}

\affil[5]{\orgname{University College London}, \orgaddress{\city{London}, \postcode{WC1E7JE}, \country{UK}}}


\abstract{The capacity-maximization design philosophy has driven the growth of wireless networks for decades. However, with the slowdown in recent data traffic demand, the mobile industry can no longer rely solely on communication services to sustain development. In response, Integrated Sensing and Communications (ISAC) has emerged as a transformative solution, embedding sensing capabilities into communication networks to enable multifunctional wireless systems. This paradigm shift expands the role of networks from sole data transmission to versatile platforms supporting diverse applications. In this review, we provide a bird's-eye view of ISAC for new researchers, highlighting key challenges, opportunities, and application scenarios to guide future exploration in this field.}




\maketitle
{\large \textbf{Key points}}
\begin{itemize}
  \item Current data traffic demand is slowing down, the mobile industry is looking for other economic drivers rather than solely relying on communication economic returns.
   \item Integrating wireless sensing functionality into the existing wireless network requires a slight modification to the hardware and signal processing pipeline. 
    \item Perceptive networks trigger three fundamental shifts: from device-based localization to ubiquitous wireless sensing, from monostatic to collaborative multistatic sensing architectures, and from joint signal processing to dedicated sensing data transmission.
    \item Perceptive network addresses the emerging need for the low-altitude wireless network, in which sensing and communication capabilities serve as the transportation infrastructure to manage low-altitude airspace traffic.
\end{itemize}

\section{Introduction}\label{sec1}

\begin{figure*}[t]
\centering
\includegraphics[width=1\textwidth]{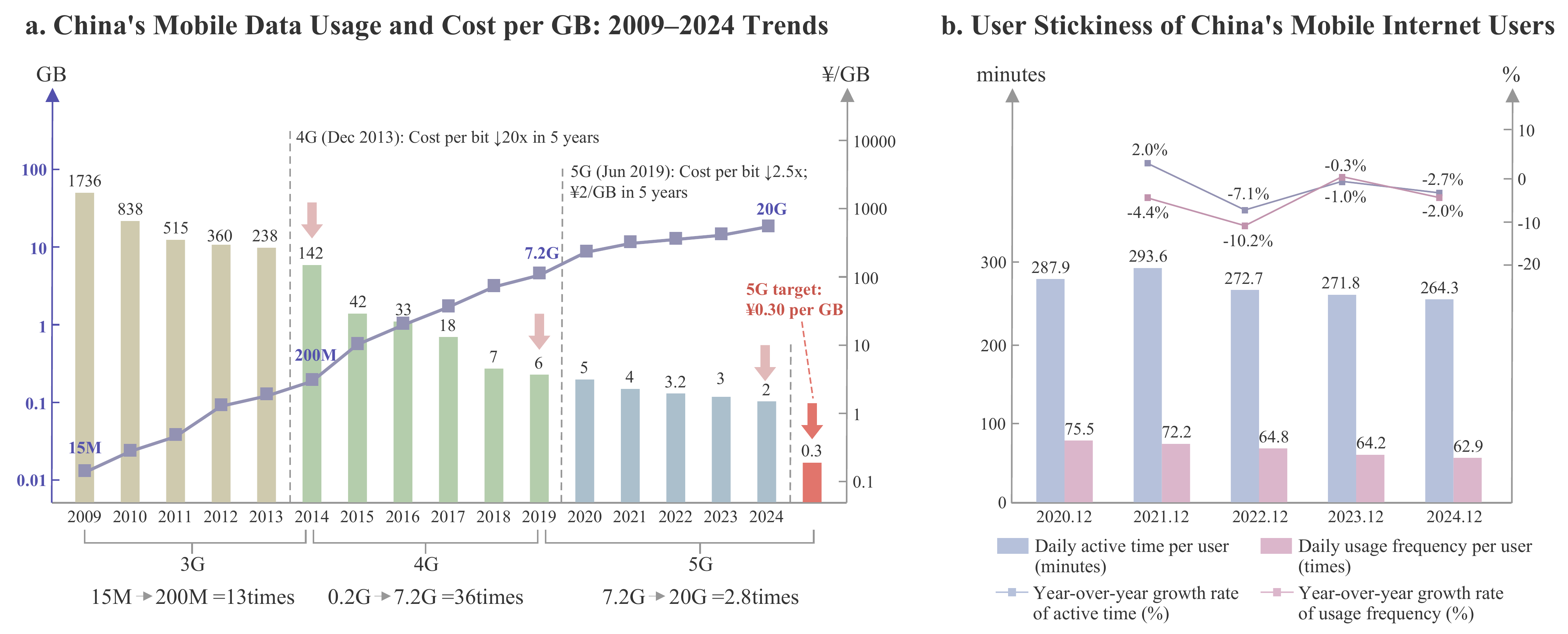}
\caption{\textbf{Usage–cost divergence and user engagement stagnation in China’s mobile internet.} (\textbf{a}), Trends in mobile data usage and unit cost in China. Following the launch of 4G in 2013, the cost per GB dropped $20$× in five years, while 5G, launched in 2019, saw only a $2.5$× reduction over a similar period, with unit costs plateauing around ¥2 per GB, far from the ¥$0.3$/GB target envisioned for 5G. This slowdown reflects weak traffic growth and stable ARPU, breaking the long-held belief that increased capacity guarantees demand. (\textbf{b}), User stickiness of China's mobile internet users. A persistent downward trend reveals weakening user engagement, with negative year-on-year growth across most of the period.}
\label{trend}
\end{figure*}


Network capacity maximization has been a key driver of wireless network development in the past few decades and a foundational pillar in the rapid advancement of the information and communication technology (ICT) industry. Jakob Nielsen, the founder of the Nielsen Norman Group, observed that high-end users' connection speeds double every 21 months, a trend known as Nielsen's law \cite{nielsen1998nielsen}. This follows a similar but slightly slower (10\%) pace compared to Moore's law \cite{moore1975progress,moore1998cramming}, and has remained remarkably consistent to the present day. 

However, the capacity-maximization design philosophy can no longer sustain next-generation wireless network development, given the recent slowdown in data traffic demand. 
In particular, while mobile data traffic has surged exponentially since the introduction of 3G/4G, the global average revenue per user (ARPU) has plateaued or declined during the early stages of the 5G deployment \cite{Alanblog2022}, indicating that users are consuming more data without proportional payment (Fig.~\ref{trend}). Even more concerning, both daily active time and usage frequency per user declined in China, the world's largest mobile market, between 2024 and 2025. Concurrently, the growth in the per-user data traffic exhibited a marked slowdown and eventually plateaued, suggesting a saturation trend in user engagement and mobile service consumption. The simultaneous decline in data traffic and ARPU is actually shaking the pillar of the economic growth of the mobile industry, as well as Nielsen's law \cite{CharlesScience2020}.

Why is mobile data growth stalling? The answer lies in the declining innovation of high-layer applications in both scale and diversity, which directly limits the growth of daily traffic volume and their economic returns \cite{Weissberger2023Analysys}. Although wireless communication technology may be able to exploit larger bandwidth at terahertz frequencies, the associated deployment costs arising from reduced cellular coverage and heightened signal propagation losses, may outpace the short-term revenue gains of telecommunication operators \cite{GSMA2024SOMIC}. Consequently, the conventional design paradigm, which exclusively focuses on capacity maximization, may not be sustainable for the evolution of the 5G/6G network.

The diminishing of capacity-centric wireless design necessitates a fundamental reconsideration of the next-generation network design. One well-recognized approach is Integrated Sensing and Communications (ISAC), which transforms wireless infrastructures into multifunctional systems with inherent sensing capabilities, unlocking new economic value. Nowadays, the International Telecommunication Union (ITU) has recognized ISAC as one of the six core usage scenarios for 6G, highlighting its critical role in environment reconstruction, target detection, and pose estimation \cite{itu2022m2516}.

In the past two decades, sensing and communication technologies have followed largely independent developmental trajectories, as sensing systems rely on radar \cite{HaschTMTT2012}, LiDAR \cite{WuSensor2021}, and cameras \cite{janai2020computer} to achieve precise environmental modeling \cite{CampbellISSC2018}, while communication systems focus on high-speed, low-latency data transmission \cite{goldsmith2005wireless,AndrewsJSAC2014,3gpp38913,TatariaProceeding2021}. However, increasing spectrum scarcity, rising hardware costs, and the growing demand for real-time sensing and high-throughput communication in emerging applications such as autonomous driving and unmanned aerial vehicle (UAV) swarms have already made ISAC an inevitable direction for future networks \cite{LiuJSAC2022,ZhangComST2022}.

In this Review, we discuss how recent advances in ISAC can serve as a transformative driver for the wireless communications industry, particularly in light of diminishing economic returns from traditional mobile data services. Considering that the cellular network has been the largest sensor network throughout human history, we believe that the opportunities of the next-generation network can be found outside of the purpose of information transmission, which we refer to as the multifunctional network. This paradigm will transform cellular and WiFi networks into the most extensive sensor network in human history. Such evolution presents significant opportunities to expand next-generation networks beyond traditional information transmission, ushering in an era of multifunctional perceptive networks. To this end, we commence on analyzing the evolution trend of two main wireless networks: the wide area network and the local area network, and introduce the technical gap when sensing is integrated into wireless communication systems. We then present our vision for the service, architecture, and networking of the perceptive network, illustrating how the multifunctional network will operate. We also detail three major application scenarios, including low-altitude airspace, automotive vehicles, and daily living scenarios, to show where the multifunctional network will serve. Finally, we identify several fundamental challenges and emerging solutions in this area to motivate future research.

\subsection{Evolution of Wide Area Network: Cellular Network}\label{subsec1-1}
The telecommunications industry plans mobile service evolution around decade-long cycles, from 1G to 6G. The ultimate success of this approach, however, has consistently hinged on striking a viable trade-off between technical ambition and economic viability. At 5G's inception in 2014 \cite{itu2015imt}, transformative applications, such as immersive virtual reality (VR), ultra-high-definition telepresence, and real-time digital twins, were envisioned as key data drivers \cite{ZhangCmag2017}. Consequently, 5G networks were designed for high capacity transmission using millimeter wave (mmWave) and massive multiple input multiple output (MIMO) technologies, targeting potential last-mile high-throughput demands between user equipment (UE) and base station (BS) \cite{RappaportAccess2013, RappaportAccess2019}. Yet a decade later, commercial penetration with mmWave remains below 1\% \cite{SambhwaniWCom2025}, with the majority of UE still operating in the sub-6 GHz spectrum \cite{ZhangVTM2019}. While operators’ reluctance to invest in costly upgrades appears responsible, the root cause is the deceleration of UE data demand, now far outpaced by system capacity \cite{gsma2021economics}. In practice, common services such as video streaming, social networks, and navigation rarely require Gbps-level data rates, rendering much of the promised 5G bandwidth underexploited in real-world deployments \cite{YuanNature2025}. 


This points to a deeper structural constraint: while cellular networks are optimized for last-hop data access, their physical-layer innovations have not yielded a commensurate increase in end-user in-service value \cite{ChenJSAC2023}. Consequently, the evolution of wireless systems, particularly 5G-Advanced and 6G networks, may shift the focus from sheer capacity enhancement to the expansion of functional capabilities\cite {ericsson2024}. A compelling alternative is to reimagine the wireless infrastructure as the world’s largest distributed sensing platform, not merely a data transmission pipeline \cite{LiuJSAC2022, PrelcicPro2024,LiuTWC2018, PangACM2024}. With BSs already densely deployed, many UEs located within 500 meters, they collectively form a perceptive network of radio frequency (RF) sensors embedded in the urban fabric \cite{LiuJSAC2022,PrelcicPro2024,LiIotJ2024}. Unlike edge servers, which remain confined to computation, BSs inherently combine proximity, power, and real-time connectivity, making them ideally suited to offer low-latency perceptive services \cite{ZhangVTM2021,CuiCMAG2025}. Viewed through this lens, 6G represents not just a leap in transmission speed, but a fundamental shift in function: from transmitting data to interpreting the physical world \cite{RongWCom2021,WangComST2023}.

Standardization of sensing capabilities in 5G-A, such as submeter positioning and static target detection using communication waveforms, marks the beginning of this transition \cite{ZhangComST2022,AlwisOJCS2021}. However, these developments remain largely incremental, realized as overlay functionalities appended to an inherently communication-centric architecture \cite{DongJSAC2025}, rather than through a fundamental reconfiguration of system design \cite{StrinatiCMAG2021, SaadNetwork2020,MuVTM2024,ZhangNetwork2024}. The vision for 6G is one of native integration, wherein sensing, communication, and computation converge seamlessly across the physical, network, and application layers \cite{WenSISCC2024,LiCMAG2025,WenIOTJ2024,WangWCom2024}. Instead of chasing ever-greater throughput, the evolution of wireless infrastructure may redefine the network from a conduit for communication into a platform for distributed perception and intelligence \cite{SaadNetwork2020,LiuSPM2023}.

\subsection{Evolution of Local Area Network: Wi-Fi Sensing}\label{subsec1-2}

Unlike wide-area networks requiring substantial infrastructure investment, which achieve economic returns primarily through high-volume data traffic, the reuse of wireless signals for sensing has gained particular traction in local area networks, notably WiFi systems. This is primarily attributed to the stagnation of traffic and data growth in Wi-Fi networks, along with their inherent advantages such as low device replacement costs and the quasi-static nature of indoor environments. Indoor applications, such as human presence detection and gesture recognition, have progressed beyond concept to practical deployment and exemplify this shift from pure data conduits to multifunctional sensing platforms \cite{HeComST2016,WangTMC2017,Chi2022WidroneW6,WangJSAC2017,WuCMAG2017}. Moreover, the finalization of the IEEE 802.11bf WLAN sensing standard in 2025 marks a transformative shift for Wi-Fi networks, integrating low-cost, privacy-conscious motion and gesture sensing into mainstream routers for home and office environments \cite{RopitaultCSM2024,DuComST2025,DuComST2025}.

 As the densest deployed, user-affordable wireless devices, Wi-Fi access points typically serve a house with several rooms. Indeed, these devices are able to be ideally positioned to evolve into an environment-aware smart home infrastructure \cite{LinTVT2025,LuoTCCN2025}. A recent large-scale deployment of Wi-Fi routers, spanning more than 10 million products, achieved $92.6\%$ accuracy in motion detection in various homes, while reducing false alarms from $63.1\%$ to $8.4\%$ and maintaining channel-state information (CSI) transmission overhead below $0.3\%$ \cite{zhu2025experiencepaperscalingwifi}. These indoor systems demonstrate how communication devices can double as environmental sensors with minimal hardware modification \cite{WangJSAC2017,TanIOTJ2022}.

Hardware limitations differentiate cellular and Wi-Fi systems when reused for sensing. Wi-Fi sensing faces inherent outdoor constraints, including restricted signal range \cite{WangJSAC2017,cisco2020wifi5g}, hardware constraints \cite{OughtonEle2023,OUGHTONTP2021}, and regulatory restrictions \cite{osenstatter2024wifi,RahmanTCCN2023,ramezanpourCN2023}. In contrast, cellular networks, particularly those operating in millimeter wave (mmWave) bands, offer significantly greater sensing potential due to their high resolution analog-to-digital converters (ADC) \cite{LiuJSAC2022,WildAccess2021,ZhangComST2022}, large-scale antenna arrays \cite{RusekSPM2013,BogaleVTM2016,XuJSTSP2025}, and precise synchronization mechanisms \cite{HuComST2021,YehExpress2023,ShresthaAccess2018}.

\subsection{Technology Gap Between Sensing and Communication}\label{subsec1-3}

From Wi-Fi sensing systems to cellular networks, the technical evolution trend of integrating sensing into wireless communications is clear. On the one hand, wireless networks will “sense what they serve”  \cite{XieWCom2023,AsabeOJVT2024,AgrawalAccess2024}, shifting from information transmission to generation. This transformation will drive large-scale innovations such as urban crowd monitoring, traffic flow estimation, and safety-critical alerts \cite{YouSCIS2021}. On the other hand, sensing-derived information will generate formidable data traffic volumes, forming the next economic pillar for wireless communications. 


Achieving seamless integration requires overcoming significant technical gaps. Current cellular waveforms, such as orthogonal frequency division multiplexing (OFDM) and orthogonal time frequency space (OTFS) offer inferior time-frequency resolution compared to dedicated radar waveforms such as frequency-modulated continuous wave (FMCW) \cite{BaqueroTMTT2019}, creating a fundamental physical layer barrier between sensing and communication \cite{LiuTSP2018,ZhengSPM2019,LiuTCom2020,ZhangJSAC2021}. Bridging this gap demands native multi-base-station cooperation and distributed sensing coordination on the user side to enable high-precision target localization and environmental perception from multiple perspectives \cite{WuJSAC2021,NematiAccess2022,MengWCom2025}. Communication systems operate over rapidly varying and stochastic wireless channels, which pose significant challenges to AI models trying to learn stable and generalizable patterns \cite{ZapponeTCom2019,IlievMIPRO2021,TongWCom2022}. 

In contrast, sensing tasks leverage spatially correlated, slowly varying environmental structures, which are highly amenable to semantic interpretation and real-time digital twin construction \cite{YangNetwork2020, ShafinWCom2020, ZhangCN2020}. As cellular networks evolve, the sensing functionalities are expected to move beyond conventional target detection, advancing toward comprehensive environmental modeling that captures static structures (e.g., walls, curbs), dynamic objects (e.g., vehicles, pedestrians), and their reflective material properties, thereby enabling more intelligent, context-aware network capabilities \cite{LiuJSAC2022,YouSCIS2021,ZhangComST2022}.

\section{Perceptive Network: How the Multifunctional Network Will Be? }\label{sec2}
For next-generation cellular network design, integrating sensing capabilities into existing infrastructures, known as ISAC technology, will drive network modifications across multiple domains, including service, architecture, and fundamental network operations. Empowered by ISAC, this network will eventually be a perceptive network. This chapter delves into these multidimensional shifts in detail.

\begin{figure*}[th]
\centering
\includegraphics[width=1\textwidth]{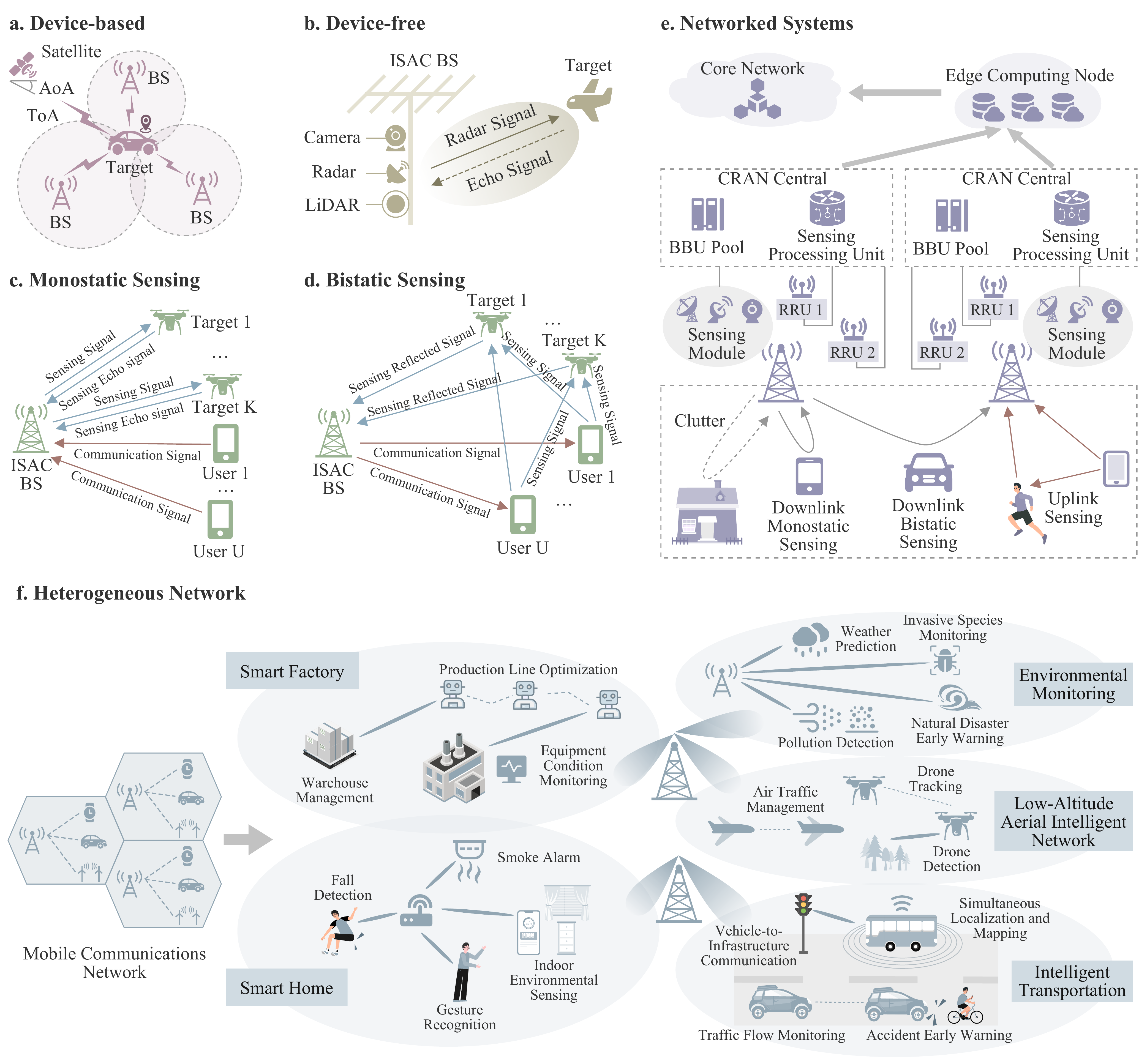}
\caption{\textbf{Illustration of the key configurations in perceptive networks.} (\textbf{a}), Device-based sensing, where the target actively transmits or receives signals and is localized via Global Positioning Systems (GPS). (\textbf{b}), Device-free sensing, where passive targets are detected through disturbances in ambient wireless signals. (\textbf{c}), Monostatic sensing with co-located transmitter and receiver. (\textbf{d}), Bistatic sensing with a spatially separated transmitter and receiver. (\textbf{e}), C-RAN system, where multiple radar-enabled RRUs connect to a centralized BBU for joint communication and sensing, with edge fusion and cloud-level optimization. (\textbf{f}), Heterogeneous perceptive network architecture integrating BSs, UEs, RSUs, UAVs, and vehicles for cooperative, multi-source sensing in shared environments.}
\label{Network}
\end{figure*}

\subsection{Service : From Device-based Localization to Device-free Sensing}\label{subsec2-1}

Current networks primarily offer device-based localization, requiring users to carry mobile devices to collect location information (Fig.~\ref{Network}a). The next generation perceptive network will enable device-free sensing, deriving not only user locations but also environmental changes, motion characteristics, and object geometries \cite{ShiJSAC2022,XiaoACMCS2022}. Beyond reusing communication signals, BSs increasingly co-locate millimeter wave radar, LiDAR, and cameras to improve environmental perception accuracy and robustness  (Fig.~\ref{Network}b) \cite{CuiJSTSP2024,XiaoACMCS2022,ShastriComST2022}. This integration formalizes the paradigm shift of the network service from device-based localization to device-free sensing.

 For example, in sensitive areas such as data centers where telecommunication devices are prohibited, BS or AP detect the presence of unauthorized persons by analyzing changes in the reflection of the wireless signal, triggering security alerts \cite{QiTGCN2023}.
 In perceptive networks, where sensing is natively embedded within the communication infrastructure, this capability transcends add-on functionality. Requiring no user cooperation, it delivers scalable, unobtrusive and continuous situational awareness, making the wireless sensing service particularly suitable for applications in smart cities \cite{ZhangTCE2024}, security \cite{YangTMC2024}, and autonomous systems \cite{huangCC2024}.

 Despite these advantages, device-free sensing still faces several limitations, including reduced localization accuracy compared to device-based methods, difficulties in distinguishing closely spaced targets, and sensitivity to environmental noise, highlighting the need for more research \cite{WangNetwork2018}.

\subsubsection{Quality of Sensing Service: Bandwidth}\label{subsubsec2-1-2}
An ISAC system must simultaneously optimize sensing and communication within limited time-frequency resources, creating an inherent bandwidth/spectrum trade-off that guides resource allocation \cite{XiongTIT2023,WangTWC2024}.
However, fundamental performance limits diverge between sensing and communication functionalities, particularly in bandwidth allocation. Typically, sensing is evaluated via the Cramér-Rao bound (CRB) \cite{LiuTSP2022,MillerTIT1978,SongTSP2023}, while communication uses Shannon capacity \cite{GuoTIT2005}. A key observation is that CRB exhibits a diminishing returns effect: sensing precision improves with bandwidth initially, but saturates beyond a threshold due to estimation errors and interference \cite{ZhangJSAC2021,XiongTIT2023}. This trade-off is further complicated by the difference between uplink and downlink sensing.

In downlink sensing, the transmitted symbols are known to the receiver, and the sensing and communication channels are correlated but distinct. In contrast, in uplink sensing, the receiver lacks prior knowledge of the transmitted symbols, and the channel remains the same, introducing additional signal estimation errors that further entangle the CRB-SINR trade-off \cite{XiongTTM2024}. Consequently, under the total bandwidth limitation, there exists an optimal bandwidth allocation point that balances communication capacity and sensing accuracy. By leveraging techniques such as joint beamforming, dynamic resource allocation, and stochastic geometric channel modeling, the CRB-SINR trade-off can be optimized to achieve a Pareto-optimal solution, thereby enhancing the spectral efficiency of ISAC systems \cite{LiuComST2022}.

\subsection{Architecture : From Monostatic Sensing to Networked Sensing}\label{subsec2-2}
Early wireless sensing research focused predominantly on monostatic architectures, where a single BS or UE managed both sensing and communication. Current networks, however, are evolving toward multi-node collaborative sensing, enabling cross-layer information fusion and dynamic resource allocation to overcome coverage limitations and enhance spatial resolution \cite{LuIoTJ2024}. 

\subsubsection{Monostatic/Bistatic Sensing}\label{subsubsec2-2-1}

Sensing architectures involving one BS and one UE are typically categorized as monostatic or bistatic systems, following radar conventions \cite{HitzlerTMTT2018}. Monostatic configurations employ shared or co-located antenna arrays for simultaneous transmission and reception, eliminating inter-node synchronization requirements \cite{LuIoTJ2024}. Monostatic systems can improve target detection accuracy by leveraging waveform diversity gain, in which multiple antennas transmit multiple independent waveforms \cite{4350230}. In next-generation perceptive networks, BSs and UEs can function as a monostatic sensing architecture, enabling uplink signals to simultaneously support data decoding and target parameter estimation (e.g., distance, angle). This achieves a seamless communication-sensing fusion (Fig.~\ref{Network}c).

Bistatic sensing, with spatially separated transmitter and receiver, increases parameter estimation complexity \cite{BrunnerTMTT2025,WuSPM2024} but enables observation of multidirectional targets and gains in spatial diversity \cite{BruyereTAES2008,PucciICCWShop2022}. Typically, BS acts as the transmitter while UE acts as the receiver, trying to extract the spatial environmental information between the BS and US. Effectively decoupling sensing echoes from downlink communication returns and reducing cross-layer interference are crucial in this setting \cite{BabuTWC2024,YuTSP2020}. Moreover, when scatterers constituting the communication channel align with sensing targets, sensing measurements could enhance the accuracy of the downlink beamforming, which is known as sensing-assisted communication technology \cite{LiuTWC2020,NieISACom2023,MengTVT2024} (Fig.~\ref{Network}d).

\subsubsection{Networked Sensing}\label{subsubsec2-2-2}
Traditional mobile networks employ cellular architectures in which BSs operate in distinct, non-overlapping coverage areas to minimize cross-cell interference\cite{AndrewsJSAC2014,BhushanCMAG2014,10726912}. The perceptive network can fundamentally disrupt this paradigm by extracting target information from interference, shifting the network from single-point coverage to a ubiquitous radio sensing network \cite{LiuJSAC2022,LiuTCom2020,DongNetwork2024} (Fig.~\ref{Network}f). In practice, various sensing entities (BSs, UEs, roadside units, unmanned aerial vehicles (UAVs), and vehicular terminals) collaborate within shared physical regions, forming a heterogeneous architecture\cite{10769538,10735119}. This provides a possibility for multi-source information fusion that enhances sensing capabilities and system robustness \cite{XieWCom2023,ZhangAPWC2017}. 

To this end, perceptive networks typically adopt the cloud radio access network (C-RAN) architecture, where multiple remote radio units (RRUs) share a baseband unit (BBU) for sensing information fusion \cite{PanCOMAG2018,ZhangVTC2027,ZhangVTM2021} (Fig.~\ref{Network}e). Cooperative RRUs simultaneously support communication and sensing, enabling both uplink and downlink sensing functions \cite{RahmanTAES2020}. As the computational core at the edge of the network, BBU not only performs traditional communication tasks such as channel estimation and modulation/demodulation, but also undertakes additional sensing-related computations, including target detection, tracking, and object recognition through cooperative sensing \cite{RahmanTAES2020}. C-RAN architecture extends this capability by enabling intercell coordination, significantly improving cross-cell target tracking accuracy. For example, crowd and vehicle flow estimation becomes feasible when geographically distributed RRUs capture real-time traffic data from multiple perspectives, fusing observations into unified mobility estimation models.

Moreover, perceptive network leverages mobile edge computing (MEC) for multi-station sensing data fusion and low-latency environmental modeling. At the core network level, a centralized cloud platform manages global sensing data storage, AI model training, and long-term optimization, enabling intelligent decision-making across large-scale networks.

\subsection{Network: From Joint Signal Processing to Sensing Data Transmission}\label{subsec2-3}
The core challenge of the ISAC signal processing architecture lies in balancing communication and sensing performance by allocating wireless resources, spanning from joint physical-layer waveform optimization to network-level scheduling and transmission. 

\subsubsection{Joint Signal Processing}\label{subsubsec2-3-1}
Sensing with communication signal via unified waveform is recognized as the most resource-efficient approach to the perceptive network \cite{LiuTSP2018,ZhangJSAC2021,WuJSAC2022,ZhouOJCS2022,DongICCC2023}. Therefore, physical layer waveform design to allocate wireless resources remains fundamental to ISAC performance optimization. On the one hand, nonoverlapped resource allocation strategies, such as time \cite{KenneyProceeding2011,KumariTVT2018,WymeerschWCom2017,ZhangJSAC2022}, frequency \cite{LiuCL2017,ShiTSP2018,KeskinTSP2021,SolaijaTVM2024}, spatial \cite{FrazerRC2007,LiuTWC2018,AndreaTWC2020}, and code division \cite{WeiTWC2008,ChenIOTJ2021,YeJSAC2022}, are practically feasible in commercial BSs, which isolate communication and sensing by constructing orthogonal resources across different domains. On the other hand, fully unified waveforms balance sensing and communication performance through sensing-centric \cite{HassanienTSP2016,BouDaherRadarConf2016,LiuCL2021}, communication-centric \cite{KeskinJSTSP2021,JohnstonJSAC2022}, and jointly optimized \cite{ChenTSP2021,ChenJSAC2022} designs. Resource allocation remains a fundamental challenge in ISAC performance optimization \cite{LiuJSAC2022,CuiNetwork2021,ChiriyathTCCN2017}. Stochastic geometry modeling of multinode sensing networks enables efficient allocation of bandwidth, power, and time-frequency resources \cite{ShenTWC2005,YanTSP2015,IslamWCom2018,ZhangTVT2020}.

From a standards perspective, current frame structures exhibit significant optimization potential for perceptive networking. Conventional uplink-downlink slot ratios (e.g.,  3:7 in 5G-Advanced frame configuration) prove suboptimal when applied directly to sensing. In time division duplexing (TDD) systems, downlink environmental echoes, whose propagation delays correlate with target distance, frequently encroach onto uplink timeslots. This interference creates annular blind zones for the detection of BSs, where echo overlap corrupts both uplink communication transmission and target detection.

\subsubsection{Sensing Data Processing}\label{subsubsec2-3-2}

Cellular data traffic is shifting from human-generated sources to machine-generated ones, which is the same as the perceptive network. In fact, sensing data now scales from hundreds of Mbps to Gbps per second per BS. Moreover, scenario requirements, such as vehicular networks and smart factories, also call for millisecond-level latency and centimeter-level positioning accuracy \cite{DongSensor2009,ZhangComST2022}. To meet these requirements and alleviate congestion, the sensing data in the perceptive network must be pre-processed before transmission \cite{CuiCMAG2025}. In this context, edge preprocessing would be crucial, facilitating raw data preprocessing, feature extraction, and filtering, which aim to reduce transmission overhead before channel coding \cite{HuangWCL2022}. Moreover, with more powerful computing at the network edge, the perceptive network would also process various sensing data from different sensors. \cite{CuiJSTSP2024,DuTWC2023}.

\section{Where the Multifunctional Network will Serve?}\label{sec3}

\begin{figure*}[t]
\centering
\includegraphics[width=1\textwidth]{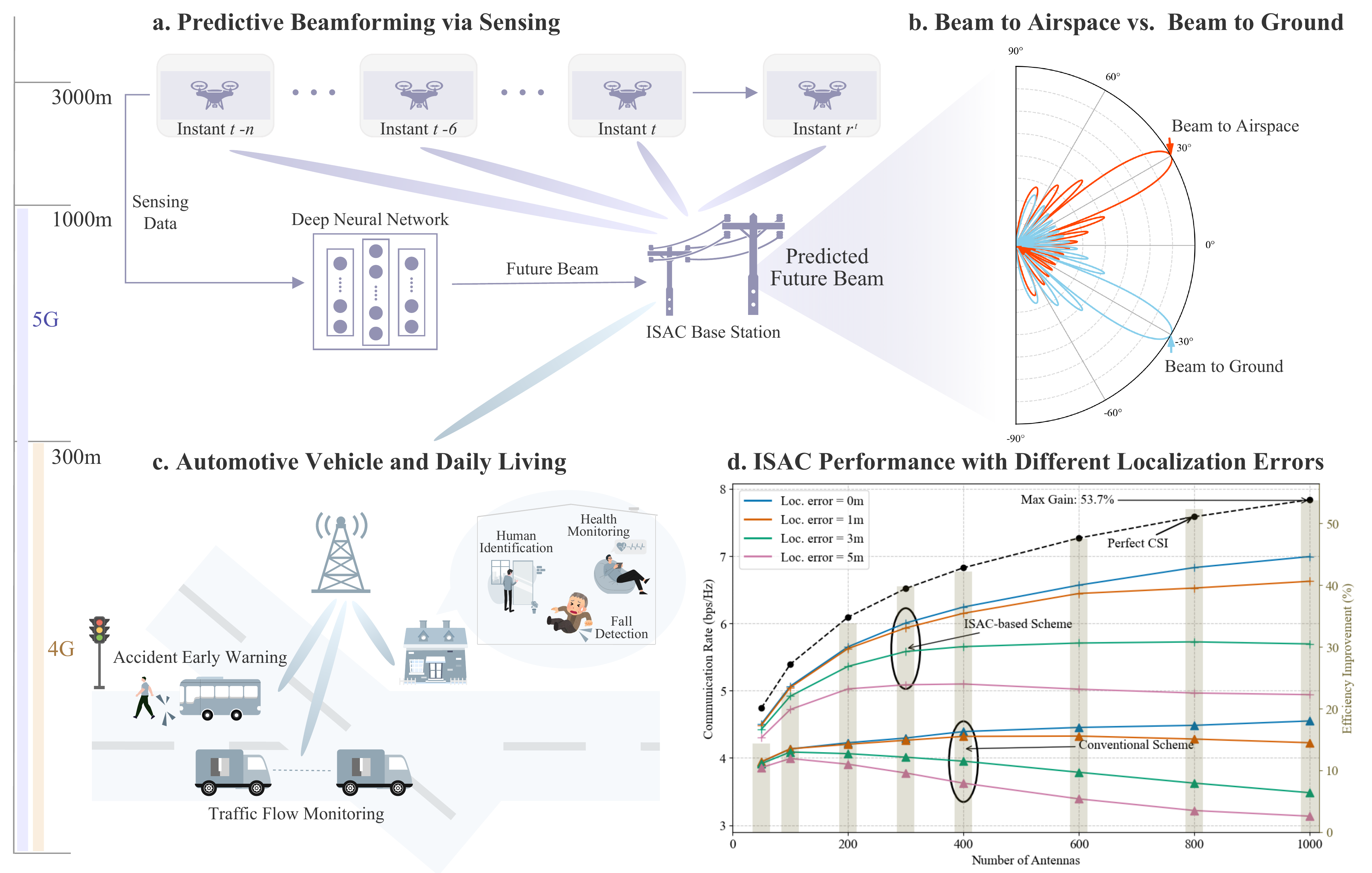}
\caption{\textbf{Overview of predictive beamforming and integrated sensing-communication performance.} (\textbf{a}), Schematic framework illustrating predictive beamforming enabled by sensing data acquisition and AI-based models, which forecast future beam directions to proactively steer beams and prevent communication interruptions. (\textbf{b}), Illustrative comparison between air-targeted and ground-targeted beamforming, highlighting the differences in beam orientation and coverage strategies for aerial versus terrestrial users. (\textbf{c}),Integration of sensing and communication in automotive and daily living environments. (\textbf{d}), Performance comparison between ISAC-based and conventional communication schemes under varying localization errors and antenna array sizes. The ISAC approach consistently outperforms the traditional baseline at localization errors of $0$m, $1$m, $3$m, and $5$m, demonstrating superior robustness to position uncertainty. The bar inset shows that the ISAC scheme achieves over $50$\% throughput gain at $0$m localization error in certain scenarios, highlighting the value of ISAC in practical multi-antenna deployments.}
\label{scenarios}
\end{figure*}

\subsection{Low-Altitude Airspace}\label{subsec3-1}
Perceptive networks address the emerging need for low-altitude wireless networks (LAWN) that support aerial vehicles, which are expected to play a central role in transforming the air transportation infrastructure \cite{HosseinIOTJ2016,hangzhou2025_lowalt}.

\subsubsection{Opportunities and Challenges }\label{subsubsec3-1-1}
Traditional wireless infrastructure remains fundamentally ground-based in design and operation, prioritizing uninterrupted human-centric communication \cite{series2015future}. This terrestrial focus overlooks the growing demand for three-dimensional services to support emerging low-altitude applications, e.g. electric vertical takeoff and landing (eVTOL) and unmanned aerial vehicle (UAV) \cite{ZhangVTM2019,WangTVT2020} (Fig.~\ref{scenarios}b). The widespread employment of UAVs and eVTOLs drives a paradigm shift in wireless infrastructure from a ground-based communication-only network to a skyward multifunctional network, known as LAWN \cite{yuan2025groundskyarchitecturesapplications}. Unlike conventional systems, LAWN functions as the transportation infrastructure to provide a unified operational framework that combines communication, sensing, navigation, and air traffic management (ATM) \cite{DongNetwork2024,JinTMC2025}. This transforms networks from passive utilities into dynamic enablers of a digitalized three-dimensional `skyway', which is mainly supported by the perceptive network\cite{MotlaghCMAG2017}.

Due to the high mobility of aircraft, as well as
terrain and weather factors, closing this gap demands
between the existing cellular network and the LAWN
requires dual-layer innovations. At the physical layer, ISAC enables BSs to merge environmental monitoring with data transmission to track flying vehicles while maintaining connectivity \cite{JiangCMAG2025}. As such, ISAC will provide core support in UAV tracking and trajectory optimization, as well as environmental awareness \cite{CuiNetwork2021}.

At the higher layer, integrating ATM functionality into wireless networks can dynamically manage airspace, enforce conflict resolution, and facilitate air traffic navigation, ultimately transforming the communication-only network into an air transportation governance system. In particular, the deployment of ISAC is expected to accelerate advancements in predictive beamforming and the construction of electromagnetic maps, laying the foundation for proactive high-reliability aerial communication systems. One of these approaches is predictive beamforming.

\subsubsection{Predictive Beamforming via Sensing}\label{subsubsec3-1-2}
By employing sensory data from the surrounding environment as prior information, one can eliminate the need for frequent channel updates and uplink/downlink feedback. In high-mobility low-altitude environments, conventional pilot-based beamforming faces critical link stability and handover latency challenges \cite{CuiJSTSP2024}. However, the dominant propagation of line of sight (LoS) and the predictable trajectories of aerial vehicles make predictive beamforming an efficient solution for air traffic management (ATM) systems\cite{GalkinGlobleCom2017} (Fig.~\ref{scenarios}a). A general beamforming process can be formulated as:

\begin{align}
\mathbf{h}\left( t+\Delta t \right) &=\mathcal{F}\left( \left\{ \mathbf{o}\left( \tau \right) \right\} _{\tau \leq t} \right)  \nonumber \\
\mathbf{w}^{\star}\left( t+\Delta t \right) &=\text{arg}\max_{\mathbf{w}}\left| \mathbf{h}\left( t+\Delta t \right) ^{\text{H}}\mathbf{w} \right|^2
 \label{eq1}
\end{align}
where $\{\mathbf{o}\left( \tau \right) \}_{\tau \leq t}$ denotes the sequence of contextual observations related to the state or environment of the transceiver, $\mathbf{h}\left( t+\Delta t \right)$ is the predicted future channel state and $\mathbf{w}^{\star}\left( t+\Delta t \right)$ represents the optimal beamforming vector preconfigured to maximize signal gain. The function $\mathcal{F} (\cdot)$ represents predictive algorithms to anticipate channel variations, enabling beam pre-alignment and pre-track that counteract signal attenuation induced by both propagation effects and device mobility \cite{YuanWCL2020} (Fig.~\ref{scenarios}a). This approach significantly enhances the stability and ultra-low-latency performance of airborne communications by preconfiguring the link before the aircraft enters the beam's main lobe. 

For instance, let us consider an aircraft approach phase, which refers to the final stage of flight as the aircraft descends toward the runway, typically below 3,000 feet (about 900 meters). This phase is critical for safe landing and demands high-precision localization, obstacle awareness, and trajectory stability. Traditional radar sensing, constrained by high costs and limited deployment, is outperformed by leveraging collaborative sensing across multiple BSs within existing perceptive networks, enabling accurate trajectory prediction and predictive beamforming \cite{YuanWCL2020}. With such a precise prediction, BSs can proactively steer directional beams toward the anticipated aircraft location, ensuring stable high-throughput links throughout descent and landing, and meeting the stringent safety and latency requirements of terminal flight operations \cite{LiuWCL2021}.


\subsubsection{Electromagnetic Map}\label{subsubsec3-1-3}
Contemporary wireless communication protocols typically rely on physical maps for statistical channel modeling, such as free-space or Rayleigh fading models, which are overly coarse and insufficient to accurately capture the underlying channel dynamics \cite{ZengWCom2021,LiWCNC2022,ZengComST2024}. In contrast, electromagnetic maps directly reflect the channel characteristics surrounding the BS or device of interest, which are independent of transceiver activities, enabling fine-grained channel information and potentially simplifying or even eliminating the need for real-time CSI acquisition \cite{ShiNatureCE2024}. Compared to the terrestrial environment, where vehicles, pedestrians, and other dynamic elements lead to constantly varying electromagnetic conditions, the relatively static nature of the low-altitude environment requires less frequent electromagnetic updates, making it inherently well-suited for the construction of electromagnetic maps \cite{HurstProceeding2025}. The continuous generation of high-resolution data by ISAC networks, coupled with recent advancements in data storage and mining capabilities, such as deep learning and large language models, has made it possible to construct highly efficient electromagnetic maps, which were previously difficult to achieve. The strong spatial correlation and quasi-static nature of wireless links enable the mapping of electromagnetic maps to channel knowledge in areas of interest, particularly suited for addressing complex scenarios such as non-line-of-sight (NLoS) propagation and rapid mobility in low-altitude dynamic airspace. For instance, by leveraging electromagnetic maps to obtain fine-grained channel information, a BS can pre-configure narrow beams and directly steer them toward an aerial vehicle during initial access, reducing the candidate beam search from 64 directions to 8 directions \cite{XiaoComST2022}.

\subsection{Automotive Vehicle Scenario }\label{subsec3-2}

\subsubsection{Opportunities and Challenges}\label{subsubsec3-2-1}
Vehicles are already equipped with a variety of sensors to support high-precision autonomous driving. These sensors form a local network that assists the vehicle in obstacle avoidance and trajectory prediction \cite{HeathSPM2020}. Beyond individual perception, sensor data from multiple vehicles can be shared and coordinated via vehicle-to-everything (V2X) communications to achieve more comprehensive environmental modeling and localization. Perceptive networks in transportation systems benefit from more powerful sensors, advanced signal processing, abundant computational resources, and greater communication capacity. For example, a vehicle can use radar or LiDAR data to infer the environment of radio frequencies, enabling model-driven channel prediction that enhances communication reliability without incurring additional spectrum usage or hardware costs \cite{PrelcicITA2016,BianWCL2024}. Benefiting from the cross-the-corner capability of near-field vehicular communication, a vehicle equipped with mmWave sensing and V2X can detect an occluded pedestrian around a blind intersection approximately $50$m away, up to $2$ seconds before visual contact.


Due to technical limitations, realizing the full potential of vehicle networks in real-world deployments remains challenging. At the modeling level, there is still a lack of unified theoretical frameworks to accurately describe the interplay between sensing and communication, making it difficult to systematically optimize joint performance \cite{LiuTVT2022,LiXFTWC2024}. At the data transmission level, the high mobility and frequent topology changes in vehicular networks pose significant challenges to existing V2X communication protocols \cite{CuiNetwork2021}. At the data processing level, even though vehicles are equipped with advanced sensors, considerable heterogeneity in the modality and temporal misalignment across different platforms complicate multimodal data fusion and decision-making, especially in collaborative scenarios \cite{CuiJSTSP2024,NieGlobeWK2023}. We identify radio frequency (RF)-based simultaneous localization and mapping (SLAM) and clutter suppression as two critical technical challenges requiring in-depth analysis.

\subsubsection{RF based SLAM}\label{subsubsec3-2-2}
RF-based SLAM serves as a key complement to vision- or LiDAR-based methods in vehicular perceptive networks \cite{AmjadAccess2023}. RF signals can partially penetrate obstacles, enabling reliable localization in cluttered or low-visibility environments \cite{LiuJSAC2022}. ISAC signals can be leveraged in RF-based SLAM to reduce bandwidth overhead and enhance spectral efficiency. The downlink communication signals of the perceptive mobile network framework can be reused for uplink sensing, enabling real-time SLAM without additional bandwidth consumption \cite{OuyangWCL2022}. Additionally, near-field wavefront curvature provides additional spatial degrees of freedom (DoFs) \cite{NieTMC2025}, improving SLAM accuracy through multi-bounce RF echoes. Beyond robustness in complex environments, RF-based SLAM offers enhanced security compared to optical sensor-based methods. RF signals are inherently less susceptible to adversarial attacks. Secure localization schemes that combine physical layer authentication with multipath fingerprinting have been shown to effectively prevent location spoofing and malicious signal injection, significantly outperforming vision-based approaches under adversarial conditions \cite{MitevBITS2023}.

\subsubsection{Clutters Suppression}\label{subsubsec3-2-3}

Multipath components that remain nearly constant over a period of interest and exhibit near-zero Doppler shifts are typically regarded as clutter. In traditional radar systems, clutter suppression is usually performed in the delay–Doppler domain and relies on the low correlation between target and clutter signals, such as differences in velocity, range, or angle. However, in perceptive networks, this assumption does not always hold, as both desired and clutter signals may originate from similar reflectors, such as buildings or road surfaces, making discrimination challenging \cite{LuTVT2023}. More suitable approaches exploit signal correlation across time, frequency, or spatial domains,  applying recursive averaging or differential operations to construct or suppress clutter \cite{WangJSAC2024}. Recent studies have begun exploring adaptive forgetting factors, low-rank modeling, and deep learning–based background modeling techniques to more precisely separate dynamic targets from static environmental echoes, thereby enabling more robust clutter suppression in complex mobile environments \cite{RahmanTAES2020,XuICCC2024}.


\subsection{Daily Living Scenario}\label{subsec3-3}

\subsubsection{Opportunities and Challenges}\label{subsubsec3-3-1}

In daily life, human activities alter the propagation properties of the physical environment, which dynamically affect the propagation of wireless signals through refraction, reflection, and scattering. This poses significant potential for re-using wireless communication signals to sense human behavior, with the benefit of ubiquity, non-intrusive, and long-term monitoring capability. Hence, wireless sensing is able to act as a privacy-preserving alternative or complement to cameras in the daily living scenario. For instance, in hospitals or homes, radar/Wi-Fi systems enable early detection of Parkinson's disease through long-term monitoring of a human's specified behavior. Similarly, a well-deployed BS could enable millimeter-scale long-term deformation monitoring of historical buildings via echo accumulation analysis. 

Despite these advantages, re-using communication signals for sensing still faces fundamental limitations. Firstly, in the Wi-Fi system, WiFi bandwidth constraints (20-160 MHz) yield range resolution around 1.5-15 meters, which prevents wireless sensing systems from RF imaging and multi-target detection in dense environments. More critically, while human motion primarily modulates signal phase, bistatic configurations (e.g., Wi-Fi router to smartphone links) actively suppress these variations through automatic clock synchronization, employing phase-locked loops (PLLs) and automatic frequency control (AFC) that inherently cancel motion-induced phase shifts. Finally, current wireless recognition systems rely excessively on micro-Doppler signatures that are generated from the Fourier transform. We will discuss the current advances that may be an alternative way to perform unified signal representation for both communication and sensing in the following section.

\subsubsection{Delay-Doppler Processing}\label{subsubsec3-3-2}

Traditionally, in wireless sensing systems, delay and Doppler responses, i.e., the delay-Doppler pair $(\tau, v)$ associated with the reflected echo, are used as primary coordinates or indices for characterizing the sensing echo. In other words, the sensing system directly captures the physical parameters $\eta$ of the reflectors/scatters in the DD plane, while ignoring the time-frequency related components in the channel information. In this sense, the objective of sensing is \textit{almost} identical to the DD domain channel estimation, where resolvable propagation paths are fully exploited and characterized for information transmission. A slight difference is that the channel estimation always pursues the propagation path with maximum channel gain, while sensing systems are interested in the characterization of a certain object.

A general channel estimation strategy can also be regarded as a target detection strategy, where a pilot symbol is placed at (0,0) and then transformed into the time-frequency domain via a 2D
inverse symplectic finite Fourier transform (ISFFT). The pilot goes through a pulse-shaping filter in the time-frequency domain before being emitted. After that, the DD channel response circularly convolves with the pilot, which is then post-processed at the receiver. The exact localization of the 2D peak, including the integer and fractional components of the delay and Doppler parameters, is the to-be-estimated channel parameters to generate a delay-Doppler image. Obtaining a high-resolution delay-Doppler image of the illuminated environment is a shared goal of wireless sensing and OTFS channel estimation in DD processing. As stated in the literature, the fixed waveform faces the barrier of the Heisenberg uncertainty principle, where increasing delay resolution will decrease the Doppler resolution, and vice versa. 

A \textit{well-designed} waveform/pulse is the key tool to combat this bottleneck. When the received signal goes through a matched filter conducted based on the transmitted waveform, multiple replicas of the transmitted waveform, delayed by $\tau$ and frequency shifted by $v$, are coherently combined to obtain a high-resolution result. In fact, this procedure is equivalent to the high-resolution imaging problem in the optical system, where a high-resolution image is the convolutional output of multiple low-resolution images with a specified point spread function as the kernel \cite{guey1998diversity}. It is well known that the ideal transmitted waveform (convolution kernel) is a Dirac delta function located at point (0,0) of the DD plane, which is, however, practically unprocurable. The resolutions of sensed delay-Doppler image of the illuminated environment are jointly determined by the transmitted waveform and the system setting, such as bandwidth $W$ and the time duration $T$.

\begin{figure*}[t]
	\centering
	\includegraphics[width=\textwidth]{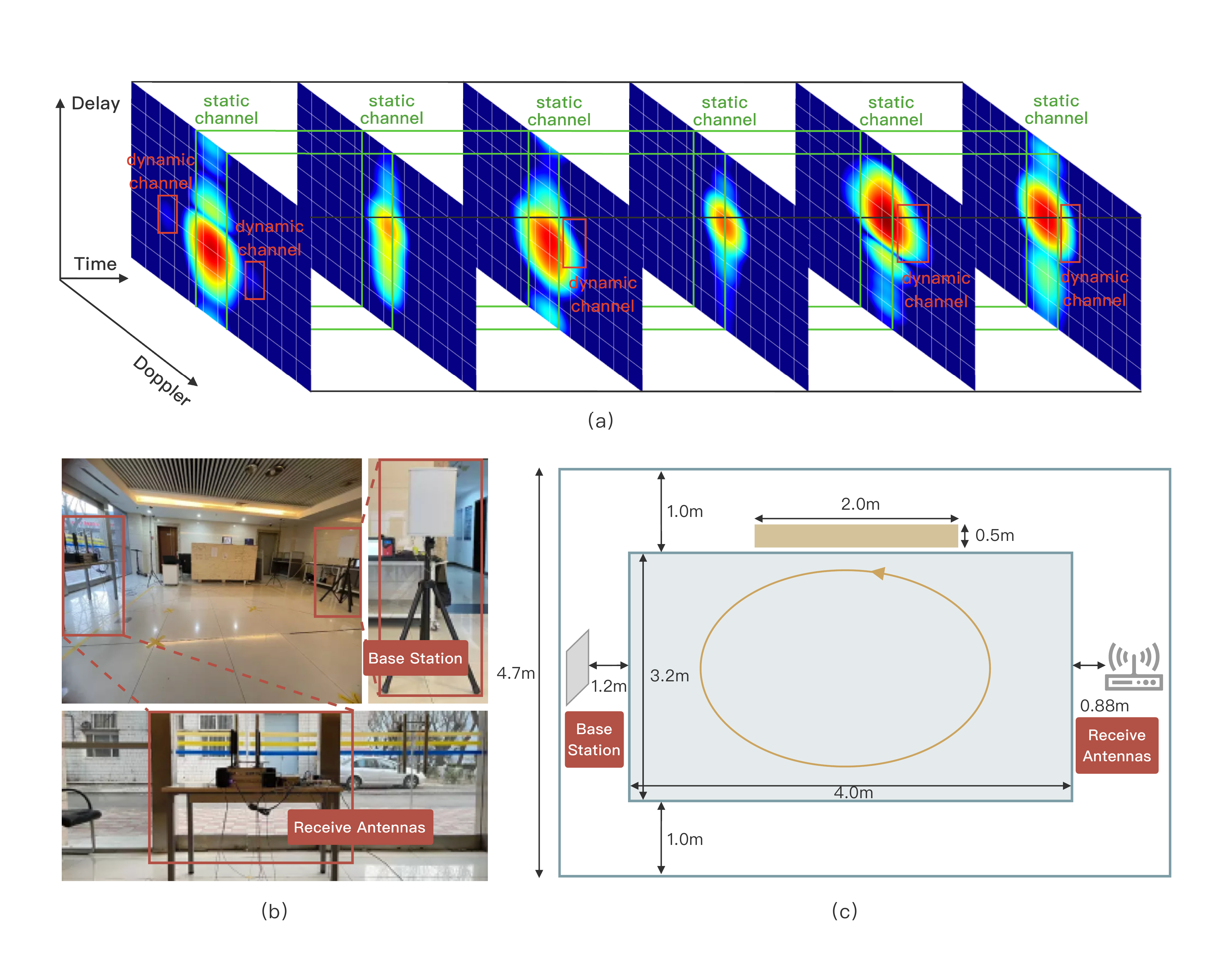}
	\caption{The results of an indoor experiment using OTFS processing, where a 5G BS and receive antennas are operating at 3.6 GHz. The subcarrier spacing is 30KHz and the carrier bandwidth is 100MHz. A person runs along an oval in the field, as shown in (c). The DD channel responses (the DD image) are shown in (a).}
	\label{Fig3}
	\end{figure*}

To further investigate the characteristics of the channel response in the DD domain, an experiment is carried out in an indoor scenario of 28 m$^2$, where a volunteer slowly runs along an oval as shown in Fig.~\ref{Fig3} (c). The BS is emitting a standardized new radio waveform (NR) at 3.6 GHz, with a $30$ KHz subcarrier spacing and a $100$ MHz carrier bandwidth. The reference signal is the reference signal for channel state information defined in the 3GPP standards. The resultant DD image is shown in Fig.~\ref{Fig3} (a), where the DD image (the channel) is arranged side by side with a time interval of $20$. In the indoor experiment, most of the DD domain channel responses are located around the zero-Doppler or nearly zero-Doppler zone because most of the objectives in the surrounding environment are static. The $100$ MHz carrier bandwidth results in an approximately $1.5$ m range resolution, therefore, the DD channel responses are cluttered over the delay dimension. Because the volunteer is running through the direct signal path between the BS and the receive antennas, the channel response reflected by the human body gradually changes over the delay dimension at different time slots.

\section{Fundamental Challenges and Emerging Solutions}\label{sec4}
The unification of sensing and communication remains fundamentally challenging, as radar and communication systems are inherently inclined to extract distinct types of information from the received signal. Specifically, radar systems are designed to exploit echo signals to minimize the uncertainty associated with the external environment, whereas communication systems seek to suppress the randomness introduced during signal transmission to reliably deliver embedded information. This fundamental divergence in signal processing paradigms brings many conflicts in ISAC system design, including the incompatibility of performance metrics, waveform design constraints, and resource allocation trade-offs.

\subsection{The Fundamental Trade-off between Randomness and Deterministic}\label{subsec4-1}

\begin{figure*}[t]
\centering
\includegraphics[width=1\textwidth]{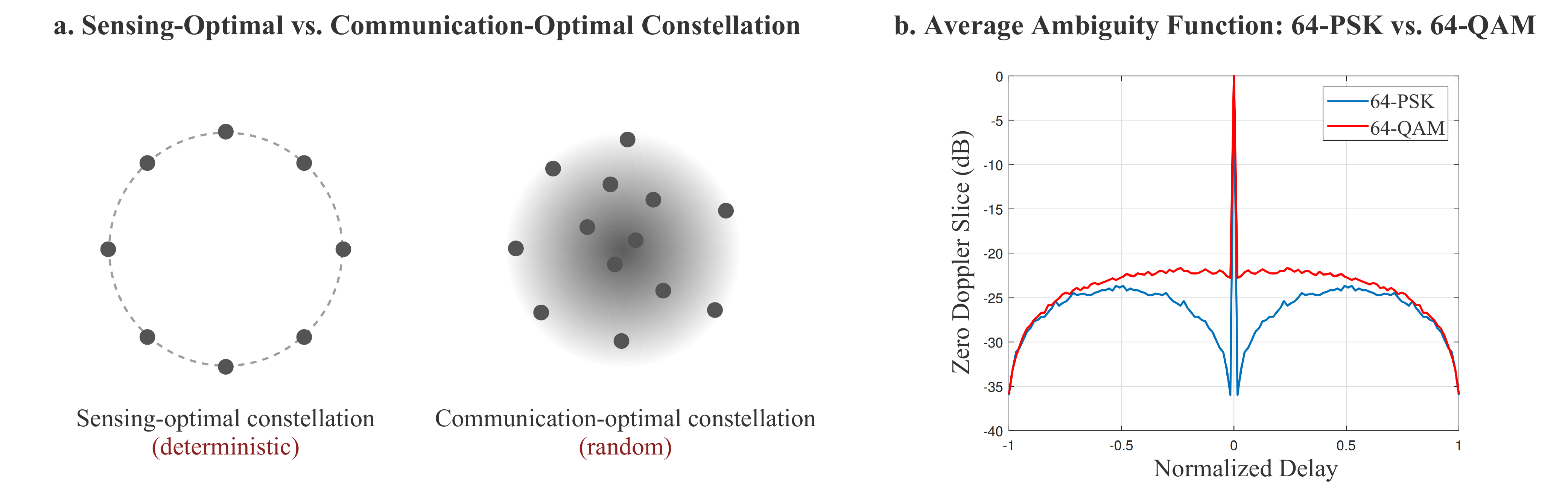}
\caption{\textbf{ Randomness–determinism tradeoff} \textbf{a,} Illustration of sensing-optimal versus communication-optimal constellations. Sensing-optimal constellations emphasize deterministic structure, while communication favors random constellations. \textbf{b,} Comparison of ambiguity functions for 64-PSK and 64-QAM. 64-PSK shows lower sidelobes, offering better delay and Doppler resolution for sensing.}
\label{random}
\end{figure*}

The aforementioned divergence is further reflected in the inherent tradeoff between randomness and determinism. Communication systems are inclined toward highly random constellations to carry more information, with Gaussian signaling theoretically proven to achieve capacity-optimal performance. In contrast, sensing performance has been shown to benefit from more deterministic constellations \cite{XiongTIT2023} (Fig.~\ref{random}a). For instance, phase shift keying (PSK), when compared to quadrature amplitude modulation (QAM), exhibits lower sidelobes in its ambiguity function, leading to more precise delay and Doppler estimation \cite{LuTSP2024} (Fig.~\ref{random}b). 

Sensing-centric waveform designs, while offering enhanced estimation performance through deterministic structures, are often incompatible with existing communication networks. As such, increasing emphasis has been placed on exploring sensing-optimal solutions under the constraints of modern communication systems \cite{DuTSP2024,LiuTSP2025}. However, to convey meaningful information, ISAC signals must inherently exhibit randomness, as purely deterministic waveforms lack the entropy required for data modulation. This randomness of the signal, while essential for communication, inevitably degrades the performance of the detection by introducing fluctuations in ambiguity and estimation metrics. Therefore, it becomes imperative to explore the deterministic-random trade-off to jointly optimize sensing fidelity and communication efficiency within a unified signaling framework.

We consider a canonical monostatic MIMO ISAC BS to serve a communication user and to perform target detection or parameter estimation for one or multiple sensing targets. The signal models for the communication and sensing subsystems follow a generic linear Gaussian formulation:
\begin{equation}
\begin{aligned}
\mathbf{Y}_c &= \mathbf{H}_c \mathbf{W}\mathbf{X} + \mathbf{Z}_c \\
\mathbf{Y}_s &= \mathbf{H}_s \mathbf{W}\mathbf{X} + \mathbf{Z}_s
\end{aligned}
\end{equation}
where $\mathbf{X}$ denotes the transmitted signal matrix; $\mathbf{H}_c$ and $\mathbf{H}_s $ represent communication and sensing channels, respectively; $\mathbf{Z}_c$ and $\mathbf{Z}_s$ are additive white Gaussian noise (AWGN) matrices in the communication and sensing receivers; $\mathbf{Y}_c$ and $\mathbf{Y}_s$ denote the corresponding received signal matrices. $\mathbf{W} $ is the precoding matrix to be optimized. The communication performance is quantified via the point-to-point MIMO achievable rate:
\begin{equation}
R(\mathbf{W}) = \log_2 \det \left( \mathbf{I} + \frac{1}{\sigma_c^2} \mathbf{H}_c \mathbf{W} \mathbf{W}^H \mathbf{H}_c^H \right)
\end{equation}
where $\sigma_c^2$ denotes the noise power at the communication receiver. On the sensing side, a loss function $f(\mathbf{W}; \mathbf{X})$ is defined to characterize radar performance, such as estimation error, detection probability, or CRB. The ISAC optimization problem is thus formulated as a constrained stochastic optimization problem:
\begin{equation}
\begin{aligned}
\min_{\mathbf{W}} \quad & \mathbb{E}_{\mathbf{X}}[f(\mathbf{W}; \mathbf{X})] \\
\text{s.t.} \quad & \|\mathbf{W}\|_F^2 \leq P \\
\quad & R(\mathbf{W}) \geq R_0 \
\end{aligned}
\end{equation}
where $P$ represents the power constraint imposed on $\mathbf{W}$, and $R_0$ denotes the minimum required communication rate. Solving the optimization problem typically involves tackling a non-convex objective under a non-linear communication constraint, which poses significant algorithmic challenges. Various approaches, such as alternating optimization, semidefinite relaxation, and stochastic approximation, have been proposed in prior works to address similar formulations. In particular, leveraging the structure of the sensing loss function, e.g. , its smoothness, convexity, or gradient form, can facilitate more efficient solution methods.

\subsection{Muti-Source Sensing}\label{subsec4-2}

Multi-source sensing in ISAC networks refers not merely to aggregating data from heterogeneous sensors, but to the coordinated sensing activities across spatially distributed ISAC transceivers. Such cooperation introduces fundamentally new degrees of freedom (DoF) in both spatial and angular domains. Compared to single-cell ISAC, cooperative networks can expand the sensing coverage and improve angular diversity, enabling high-resolution detection in wider fields of view. According to recent findings in multistatic cellular ISAC systems \cite{LiTWC2024}, each ISAC BS can receive not only the echo from its own transmission but also the target reflections of neighboring BSs or uplink users, forming a virtual multi-static radar topology. This allows the network to extract more granular features of the environment while reducing blind spots, much like a city watched from multiple security cameras at once, each from a different angle. On the communication side, coordinated multipoint (CoMP) transmission techniques enable multiple ISAC transmitters to serve users jointly, suppressing inter-cell interference and improving throughput \cite{HosseiniTSP2016}. In essence, cooperative ISAC networks transform isolated sensing and communication processes into a spatially integrated system capable of achieving joint performance gains.

Yet, the benefit of collaboration comes at a price, primarily in the form of coordination overhead and topological complexity. Multisource sensing requires the exchange of diverse and often delay-sensitive information such as raw echoes, environmental CSI, semantic-level features, and even neural network gradients for learning-based inference. These exchanges must be timely and synchronized, especially under mobility and environmental dynamics, which significantly stress the fronthaul and control plane capacity of current cellular architectures. Observations of the same target acquired from multiple vantage points must first be transformed into a unified coordinate system by precise spatial registration. With such an alignment, spatial redundancy can be effectively mitigated. For example, location priors can be exploited to suppress noninformative signal components during the preprocessing stage, thereby reducing transmission overhead without compromising the fidelity of the sensing outcome. However, such optimizations hinge heavily on the design of a flexible MAC layer that can support distributed inference, strict latency budgets, and adaptive spectrum sharing. As highlighted in recent ISAC network architecture studies \cite{ZhangJSAC2021}, traditional CRAN or cell-free architectures often fail to meet the timing and synchronization requirements needed for robust multi-source ISAC. Thus, a fundamental rethinking of network stack design, ranging from spectrum management to semantic-level protocol stacks, is indispensable.

\section{Outlook}\label{sec12}
The evolution of wireless communications has spurred the mobile internet industry, where sustained growth in data traffic has historically fueled 10-year network evolution cycles, from 1G to 5G. Yet, with data traffic growth stalling, the design of next-generation wireless networks urgently requires new economic drivers. This paradigm shift expands the role of networks from data transmission to versatile platforms supporting various novel applications. 

In this review, we introduced ISAC technology, which utilizes existing wireless infrastructure to provide wireless sensing services, creating a multifunctional future network with perceptive capability. We systematically analyze the evolution trends across cellular wide-area and Wi-Fi local-area networks, examining the technical gap between wireless sensing and communications. We then discussed the technical changes, including the service, architecture, and networking structure, to illustrate how the multifunctional network will be. After that, the low-altitude airspace, automotive vehicle, and daily living scenarios are extended as major application scenarios to the perceptive network, to show where the multi-functional network will serve. Finally, we identify several fundamental challenges and emerging solutions to this area.

\bibliography{sn-bibliography}

\end{document}